# MANIFESTATION OF VACANCIES IN A SPECTRUM OF ORGANIC MOLECULAR CRYSTAL LATTICE VIBRATIONS

**M.A. Korshunov**[*]

Kirensky Institute of Physics, Siberian Division, Russian Academy of Sciences, Krasnoyarsk, 660036 Russia

**Abstract**. Using a method Raman of spectroscopy and considering a non-uniformity of allocation of vacancies in a monocrystal, it is shown that lines with the small intensity (caused by presence of vacancies) have a major intensity in the sample of a larger size than in the sample of a smaller size. Increase of an intensity of lines at vacancy concentration increase can reveal which lines are related to presence of vacancies in the sample. For a p-dichlorobenzene, it is lines of a small intensity in a spectrum of the lattice oscillations in the field of 70 cm$^{-1}$.

At shootings Raman of spectrums of small frequencies is used a laser beam. And the spectrum is gained, from small field of a crystal. If flaws in particular vacancies in a crystal are proportioned nonuniformly in the given field of a crystal can be more or less flaws. Therefore the lines of a spectrum caused by the occurrence to presence of flaws, will have a different intensity depending on concentration of flaws in studied field of a crystal. Therefore an intensity of these lines from a small crystal and from a crystal of a major size will have different quantity. Thus the spectrum should be gained from all volume of the sample. In a spectrum of the lattice oscillations of a p-dichlorobenzene it is observed six intensive lines caused by orientation oscillations and a number of additional lines of the small intensity which definition is made on the basis of calculations of spectrums. In operation [1] on the basis of calculations it is shown, that lines in the field of 70 cm$^{-1}$ of a p-dichlorobenzene are caused by presence of vacancies in a crystal lattice. Calculations were spent on a method atom-atom of potentials [2]. Histograms of frequency spectrums are gained on a method the Dyne [3].

For each sample the density has been measured. To define a vacancy concentration it is possible comparing the measured density of a crystal with the density calculated on X-ray diffraction data. On the average for the studied sample the vacancy concentration has made 3-5 %.

---

[*] mkor@iph.krasn.ru

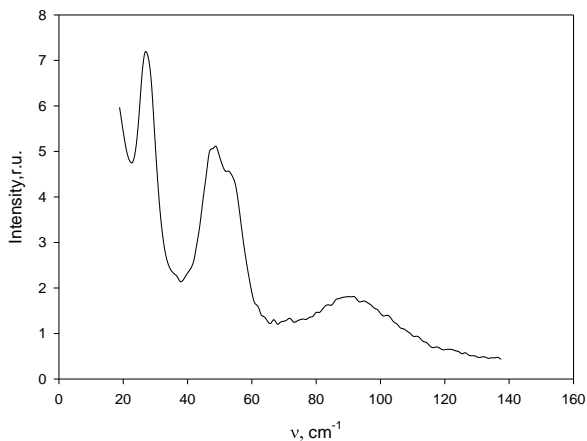

Fig. 1a.

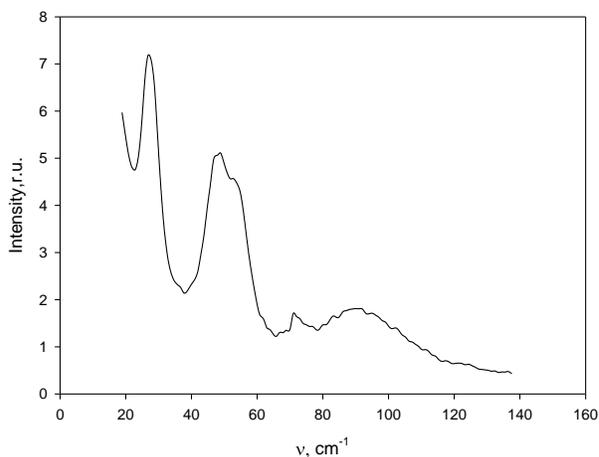

Fig. 1b.

In Figure 1a the spectrum of the lattice oscillations of a p-dichlorobenzene gained for the sample by a size of 0.1*0.2*0.3 cm$^3$ is given.

In Figure 1b the spectrum from the sample by a size of 1.0*1.0*5.0 cm$^3$ is given. As we see an intensity of additional lines of a small intensity it was incremented. In particular a line in the field of 70 cm$^{-1}$, which occurrence as show calculations it is caused by presence of vacancies. On visible from for non-uniform allocation of vacancies on volume of the sample of a line caused by the occurrence to presence of vacancies from большего the sample have a major intensity as number of fields containing vacancies more than in a small crystal.

Thus, it is possible to guess, that lines of a p-dichlorobenzene in the field of 70 cm$^{-1}$ are caused by presence of vacancies. It proves to be true the observational effects in comparison of spectrums of the lattice oscillations from samples of a various size.